\documentclass[aps,prd,twocolumn,floatfix,preprintnumbers,superscriptaddress,nofootinbib,nameinlink,capitalise]{revtex4-2}
\usepackage{style,slashed}
\usepackage[dvipsnames]{xcolor}
\usepackage{graphicx} % Required for inserting images

\begin{document}

\title{Supersymmetric Hybrid Inflation in light of Atacama Cosmology Telescope Data Release 6, Planck 2018 and LB-BK18}
\author{Mansoor Ur Rehman
\orcidlink{0000-0002-1780-1571}}
\email{mansoor@qau.edu.pk}
\affiliation{Department of Physics, Faculty of Science, Islamic University of Madinah, 42351 Madinah, Saudi Arabia}
\author{Qaisar Shafi}
\affiliation{Bartol Research Institute, Department of Physics and Astronomy,
University of Delaware, Newark, DE 19716, USA}
\noaffiliation

%\author{Mansoor Ur Rehman and Qaisar Shafi}

%\date{today}

\begin{abstract}
Supersymmetry based hybrid inflation models (called `spontaneously broken supersymmetry' by the Planck collaboration) are attractive for a number of reasons including the nice feature that inflation is associated with a local gauge symmetry breaking in the early universe. Models based on a minimal superpotential and a canonical K\"ahler potential have the important property that there is no eta problem and the supergravity corrections are adequately suppressed. Following Planck’s notation, the inflationary potential with sub-Planckian inflaton field values is approximately given by $V \simeq \Lambda^4 [1 + \alpha_h \log (\phi / M_{Pl}) ] - m_{3/2} \Lambda^2  \phi + \Lambda^4 O ( \phi / M_{Pl})^4$.
Here $\Lambda = \sqrt{\kappa} M$ denotes the energy scale of inflation, $M$ is the gauge symmetry breaking scale, $\kappa$ is a dimensionless parameter which fixes the inflaton mass ($\sqrt{2}\kappa M$), $\alpha_h$ is determined from quantum corrections in terms of $\kappa$ and the underlying gauge group, and the soft supersymmetry breaking term proportional to the gravitino mass $m_{3/2}$ ( $\sim$~10~TeV) and linear in the inflaton field $\phi$ is present during inflation. 
The final term in $V$ represents the leading supergravity correction which is well suppressed since $\phi \lesssim M_{pl}$ [Note that the last two terms were not taken into account in the Planck analysis.] We provide estimates for the parameters $\kappa$ (and $\alpha_h$ ) that yield a scalar spectral index $n_s$ in the 0.96 - 0.98 range, which is fully consistent with the recent P-ACT-LB measurements presented by the Atacama Cosmology Telescope (as well as earlier measurements by Planck.) The gauge symmetry breaking scale $M$ is determined to be on the order of $10^{15}$~GeV or so.
We recall that In the absence of the soft SUSY breaking term proportional to $m_{3/2}$ in $V$, the spectral index $n_s \simeq 1- 1/N =0.98$, where $N = 50$ denotes the number of e-foldings. The tensor-to-scalar ratio $r$ in this minimal model is tiny, but it can reach values in the observable range, $r \lesssim 0.01$, in non-minimal models.
\end{abstract}

\maketitle

\section{Introduction}

The simplest supersymmetric hybrid inflation (SHI) model (referred to by Planck as the spontaneously broken SUSY, denoted here as SBS) based on a minimal superpotential and a canonical  K\"ahler potential \cite{Dvali:1994ms,Copeland:1994vg} predicts \cite{Dvali:1994ms} a scalar spectral index $n_s = 1 -1/N = 0.98$, where $N$ set equal to 50 here denotes the number of e-foldings. 
The inclusion of soft supersymmetry breaking terms that are relevant during the inflationary epoch \cite{Rehman:2009nq}, but were ignored in \cite{Dvali:1994ms}, leads to a spectral index $n_s$ in the 0.96 to 0.97 range, preferred by the WMAP \cite{Hinshaw_2013} and Planck \cite{Planck:2018vyg,Planck:2018jri} observations.
It is important to note that the inclusion of supergravity corrections in this minimal model does not create the $\eta$ problem.
Lowering of the spectral index from 0.98 to 0.97-0.96 can also be achieved with the inclusion of non-minimal terms in the K\"ahler potential, as first pointed out in Ref.~\cite{Bastero-Gil:2006zpr} (See also  \cite{urRehman:2006hu}). This latter approach can be exploited to realize significantly larger values for the scalar to tensor ratio $r$, of order $10^{-2}$ or so \cite{Shafi:2010jr,Rehman:2010wm,Civiletti:2014bca}, which can be tested in the next generation of experiments.

This article is motivated by the recent results from the Atacama Cosmology Telescope \cite{ACT:2025tim} which, among other things, has revised the value of the scalar spectral index slightly upward compared to the Planck results. SHI (SBS) models are based on well-motivated extensions of the Standard Model with supersymmetry (SUSY) playing an essential role. Inflation is associated with the breaking of a local gauge symmetry such as $U(1)_{B-L}$, left-right symmetric models, as well as SUSY GUTs. Variations of the simplest SHI (SBS) models such as shifted hybrid inflation are designed to circumvent issues such as the primordial monopole problem \cite{Jeannerot:2000sv}. In this article, for simplicity, we restrict our attention to SHI (SBS) models in which the local gauge symmetry is broken at or near the end of inflation.

For the convenience of experimental collaborations, we parametrize the inflationary potential in the simplest SHI (SBS) models by following the notation of WMAP and Planck. In particular, we include the key terms that are present in SHI models but ignored ( represented as ellipses) in the Planck experimental papers.

As previously mentioned we use a minimal superpotential which is adequate to break the underlying local gauge symmetry and naturally also contains the inflaton field as a gauge singlet. With a minimal K\"ahler potential the scalar spectral index is in full agreement with the experimental observations. The tensor-to-scalar ratio $r$ in this minimal model is negligibly small ($ r \simeq 10^{-11}-10^{-12}$).
\section{SHI (SBS) Model}
The minimal supersymmetric hybrid inflation model (SHI), called spontaneously broken supersymmetry model (SBS) by the Planck collaboration, is based on the following renormalizable superpotential $W$ which is gauge invariant and respects a $U(1)$ R-symmetry:
\begin{equation}
W = \kappa S (\chi \overline{\chi} - M^2 ).
\end{equation}
The scalar components of the superfield $\chi$ and its conjugate superfield $\overline{\chi}$ acquire vacuum expectation values (VEVs) that break some gauge symmetry $G$ to $H$, with supersymmetry remaining unbroken. The gauge singlet superfield $S$ is required by supersymmetry to implement the gauge symmetry breaking at the tree level. Its scalar component, which plays the role of the inflaton in the standard analysis, has zero vacuum expectation value in the supersymmetric limit.
The gauge symmetry breaking scale $M$, it turns out, is determined by the inflationary scenario. Roughly speaking, the temperature anisotropy ($\delta T / T$) in the minimal model is proportional to $(M / M_{Pl})^2$, which translates to a symmetry-breaking scale $M \sim$ few $\times 10^{15}$~GeV. The dimensionless parameter $\kappa$ determines the mass of the inflaton.

For definiteness, we focus our attention on the simplest examples of $G$ such as $U(1)_{B-L}$. Of course, the breaking of a $U(1)$ gauge symmetry produces topologically stable cosmic strings and, if necessary, they can be inflated away without impacting our main conclusions in any significant way.

To complete the definition of the minimal model we introduce a canonical K\"ahler potential given by
\begin{equation}
 K = |S|^2 + |\chi|^2 + |\overline{\chi}|^2.  
\end{equation}
In the original discussion of this model \cite{Dvali:1994ms}, the inflationary potential employed global supersymmetry and ignored the soft supersymmetry breaking terms that arise during inflation.

The gauge symmetry $G$ is unbroken during inflation ($\chi = 0 = \overline{\chi}$) and the inflationary potential in this approximation is given by
\begin{equation} \label{DSS}
V = \Lambda^4 \left[ 1 +  \alpha_h\,F(x)  \right],
\end{equation}
where $\Lambda = \sqrt{\kappa} M$, $x \equiv s/M$, the field $s$ is a scalar component of the superfield $S$ (see below). The parameter $\alpha_h $ is given by
\begin{equation}
\alpha_h = \frac{\kappa^2 \mathcal{N}}{8 \pi^2},
\end{equation}
with $\mathcal{N}$ denoting the dimensionality of the representation of the superfields $\chi$ and $\bar{\chi}$. For example,  $\mathcal{N} = 1, \, 10, \, 16$ for $U(1)_{B-L}$ extension of the supersymmetric standard model, flipped $SU(5)$, and $SO(10)$, respectively. With $\mathcal{N} \lesssim O(10)$ and $\kappa \lesssim O(10)$, we expect $\alpha_h \lesssim O(10)$ in a typical realistic model. The one-loop correction is encoded in the function $F(x)$, given by
\begin{eqnarray}
F(x) &=&\frac{1}{4}\bigg((x^4 + 1)\ln{\frac{x^4 - 1}{x^4}} + 2 x^2 \ln{\frac{x^2 + 1}{x^2 - 1}} + \notag \\
 && 2 \ln{\frac{\kappa^2 M^2 x^2 }{Q^2}} -3\bigg),
\end{eqnarray}
where $Q$ is the renormalization scale. In the large-field regime, $s \gg M$, the potential simplifies to the logarithmic form:
\begin{equation}
V = \Lambda^4 \left[ 1 +  \alpha_h\log(\phi/M_{pl})  \right],
\end{equation}
assuming $Q = \kappa M_{Pl}/\sqrt{2}$. This potential, referred to as the SBS model, is used in the Planck collaboration's inflationary analysis. The field $\phi \equiv \sqrt{2} \, s$ used by the Planck collaboration is the canonically normalized real scalar field.
A straightforward calculation using slow roll approximation predicts a spectral index $n_s \simeq 1 - 1/N = 0.98$, for $N =50$, the number of e-foldings.

\begin{figure} 
\includegraphics[width=0.49\textwidth]{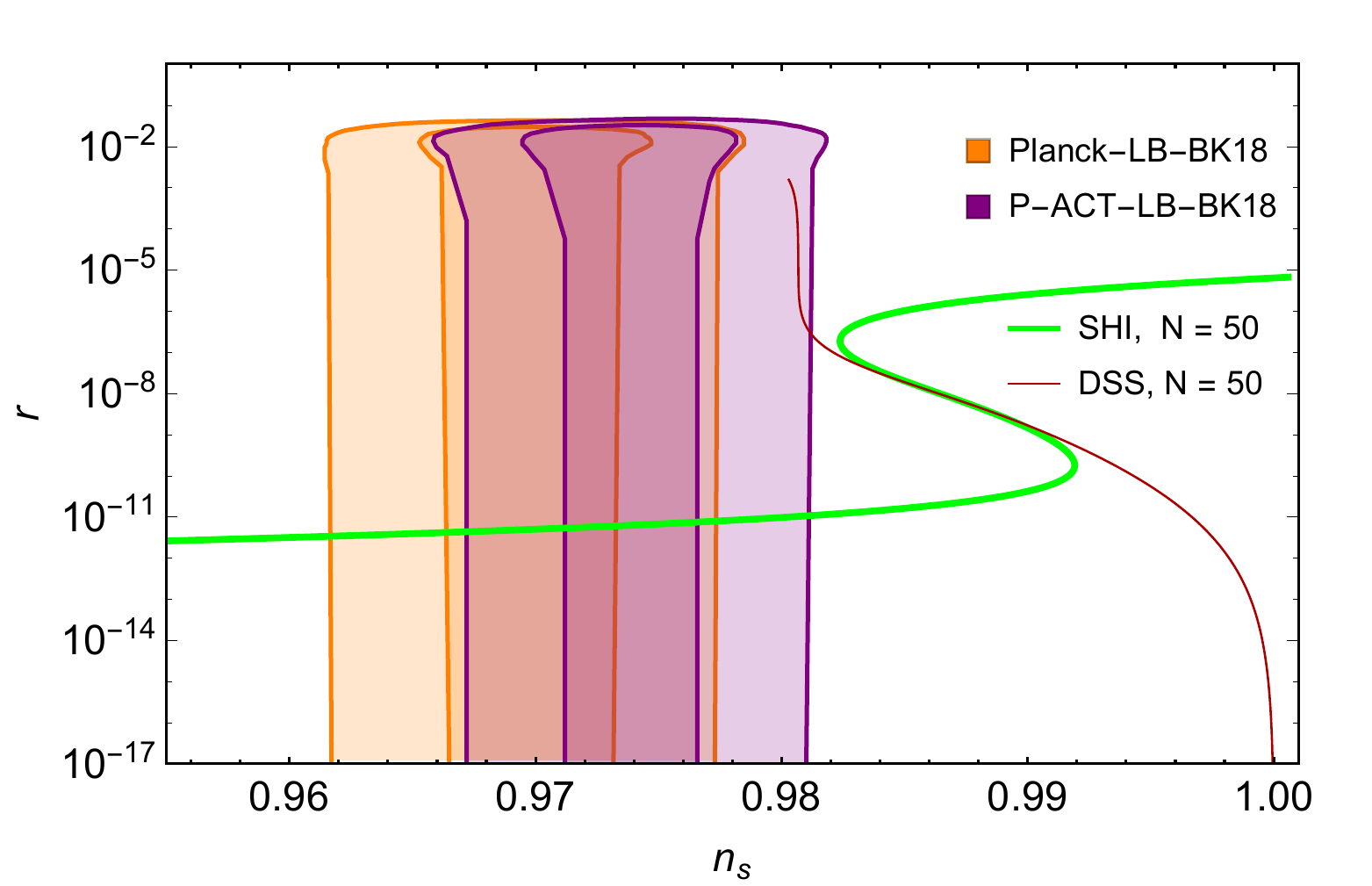}
\includegraphics[width=0.49\textwidth]{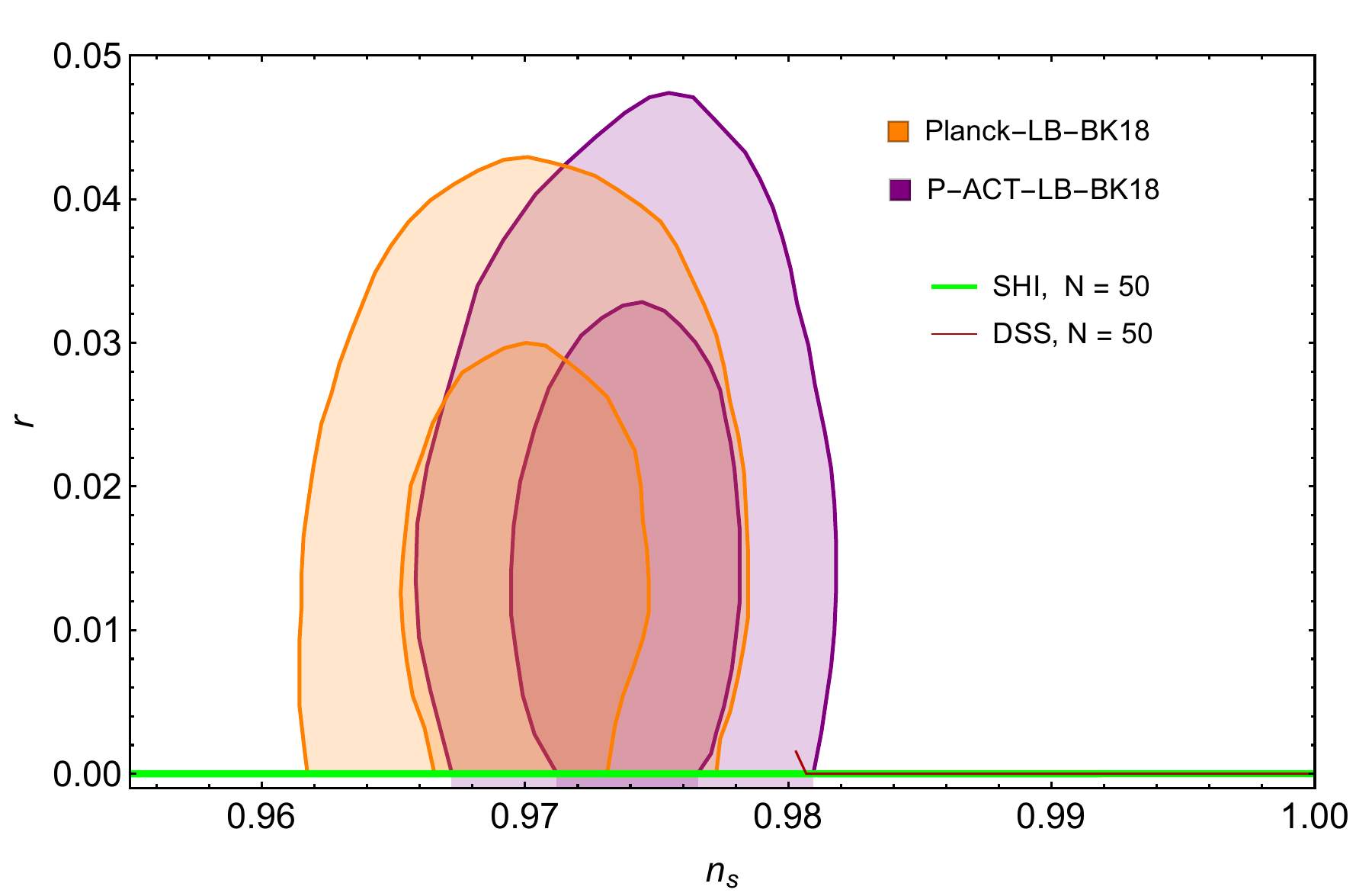}
\caption{\label{gce} $n_s$ versus $r$ predictions for SHI (thick-green curve), based on $V$ in Eq.~(\ref{vx}) (which includes soft terms and sugra corrections, and one-loop corrections), compared with data from ACT, Planck, BK18 [including LB]. In contrast, as illustrated in the DSS scenario (thin-red curve), ignoring the soft terms and supergravity corrections yields $n_s \simeq 0.98-1$. In our analysis, we set the number of e-foldings to $N = 50$.
}  \label{rns}
\end{figure}

A more complete inflationary potential takes into account the soft supersymmetry breaking terms \cite{Buchmuller:2000zm,Senoguz:2004vu,Rehman:2009nq,Buchmuller:2014epa} as well as the contributions from supergravity \cite{Linde:1997sj}.
The potential in this case is given by
\begin{eqnarray} \label{vx}
V(x)  &=& \Lambda^4 \, \bigg[1 + \alpha_h  F(x) + \frac{a \ m_{3/2} }{\Lambda^2} \phi + \frac{1}{2} \left(\frac{m_{3/2} \, \phi }{\Lambda^2} \right)^2  \notag \\
&+& \frac{1}{8}  \left(\frac{\phi}{M_{Pl}}\right)^4  \bigg],
\end{eqnarray}
with
\begin{equation}
a = 2 \sqrt{2} \  \vert 2 - A \vert \cos[\arg s + \arg (2-A)].
\end{equation}
With suitable initial conditions, the phase $\arg s$ can be stabilized before observable inflation \cite{Senoguz:2004vu,Buchmuller:2014epa}. In the following analysis, we treat $a$ as a constant of order unity ($a=-1$).
For the sake of comparison, we will refer as SHI (SBS) the model described in Eq.~(\ref{vx}), which is based on local supersymmetry and includes one-loop corrections and the supersymmetry breaking soft terms that are present during inflation. We refer to the model presented in Eq.~(\ref{DSS}) as the Dvali–Shafi–Schaefer (DSS) model, which is based on global supersymmetry and incorporates only one-loop radiative corrections.

\begin{figure}[ht]
\includegraphics[width=0.49\textwidth]{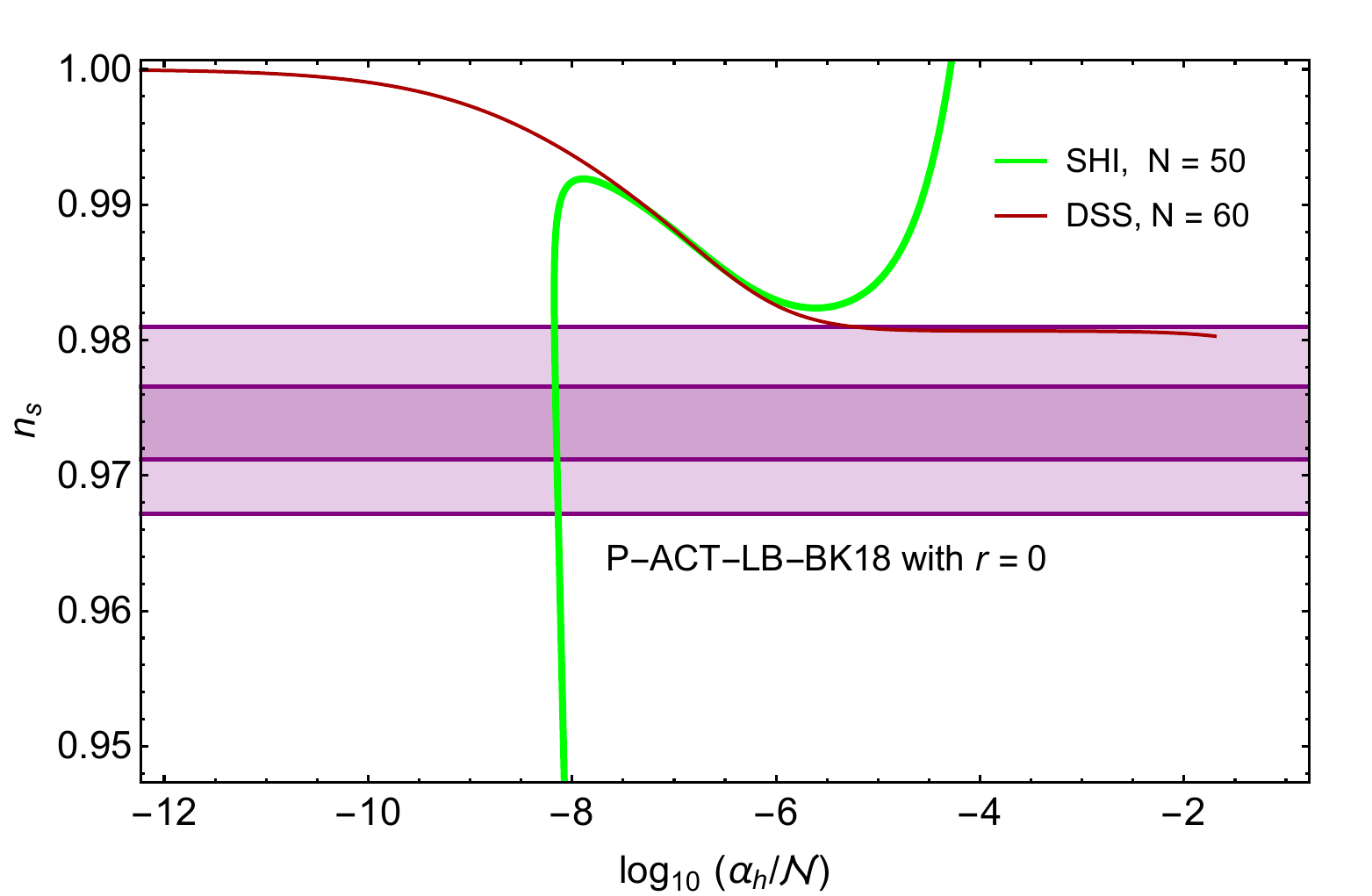}
\caption{\label{gce}$n_s$ versus $\log (\alpha_h/\mathcal{N}) = \log (\kappa^2/8\pi^2)$ for SHI (which incorporates soft SUSY-breaking terms, sugra corrections, and one-loop effects), and for DSS (with one-loop corrections only), compared with ACT, Planck, BK18 [Plus LB].
}  \label{nsa}
\end{figure}

In \cref{rns} we show two plots for $n_s$ versus $r$ based on the inflationary potential in Eq.~(7), compared with ACT, Planck, BK18 (plus LB.) The figures show excellent agreement with all existing data for $n_s$ values in the $0.96 -0.98$ range, accompanied by a tiny value for the tensor-to-scalar ratio $r$. However, in non-minimal extensions based on realistic models, the tensor-to-scalar ratio can attain values, $r \lesssim 0.01$, potentially within the reach of upcoming observations \cite{Shafi:2010jr,Rehman:2010wm,Civiletti:2014bca}.
In \cref{nsa} we show a plot of $n_s$ versus $\alpha_h$, which again shows the importance of including the soft supersymmetry breaking terms and the supergravity corrections in the inflationary potential. Note that with a gravitino mass, $m_{3/2} \sim 10$~TeV, the usual soft supersymmetry breaking scalar mass terms do not play a role in our analysis.
\begin{figure} [h]
\includegraphics[width=0.49\textwidth]{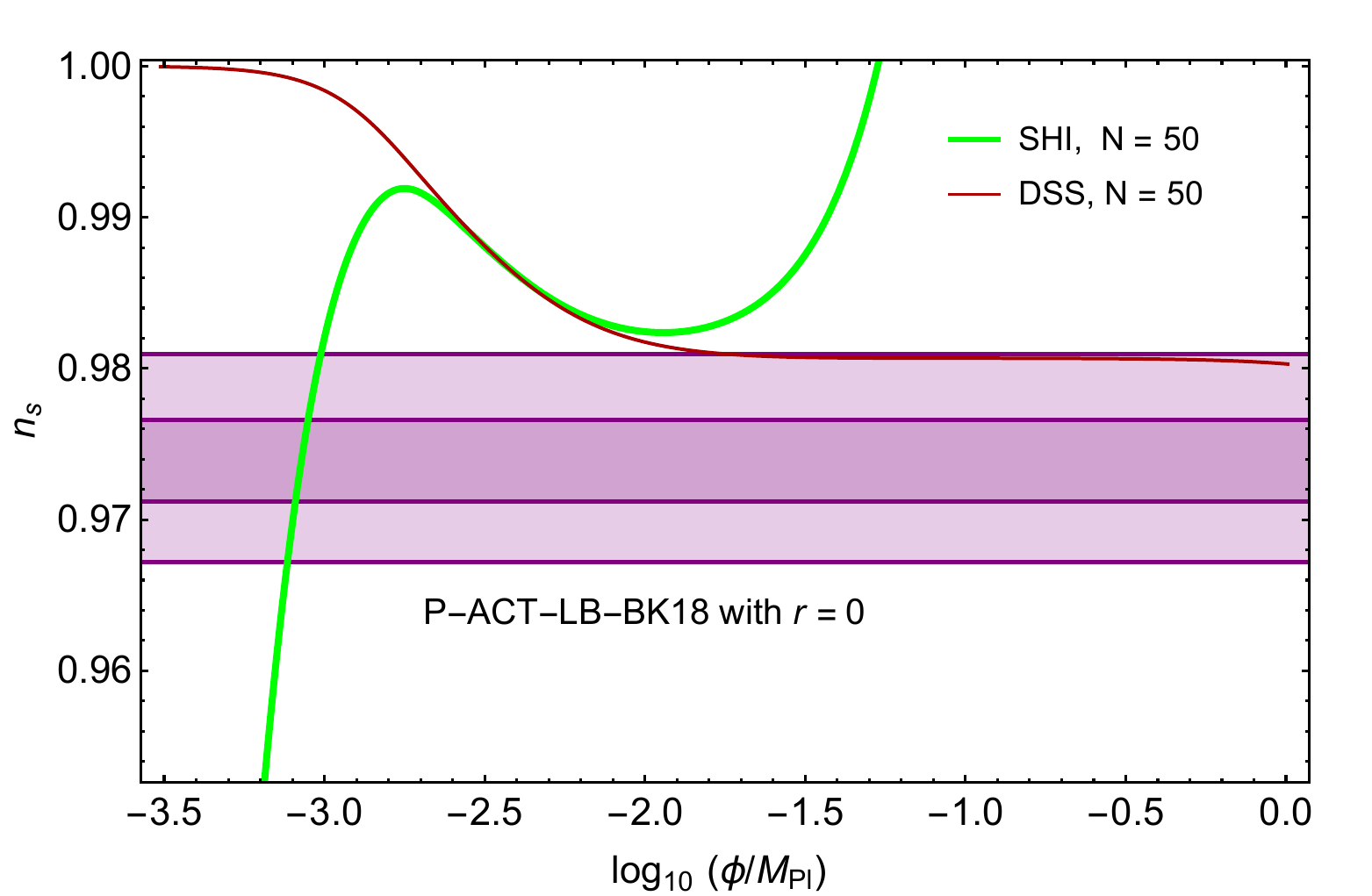}
\caption{\label{gce}$n_s$ versus $\log (\phi/M_{Pl})$ for SHI (which incorporates soft SUSY-breaking terms, sugra corrections, and one-loop effects), and for DSS (with one-loop corrections only), compared with ACT, Planck, BK18 [Plus LB].
}  \label{nsphi} 
\end{figure}
\begin{figure}[h]
\includegraphics[width=0.49\textwidth]{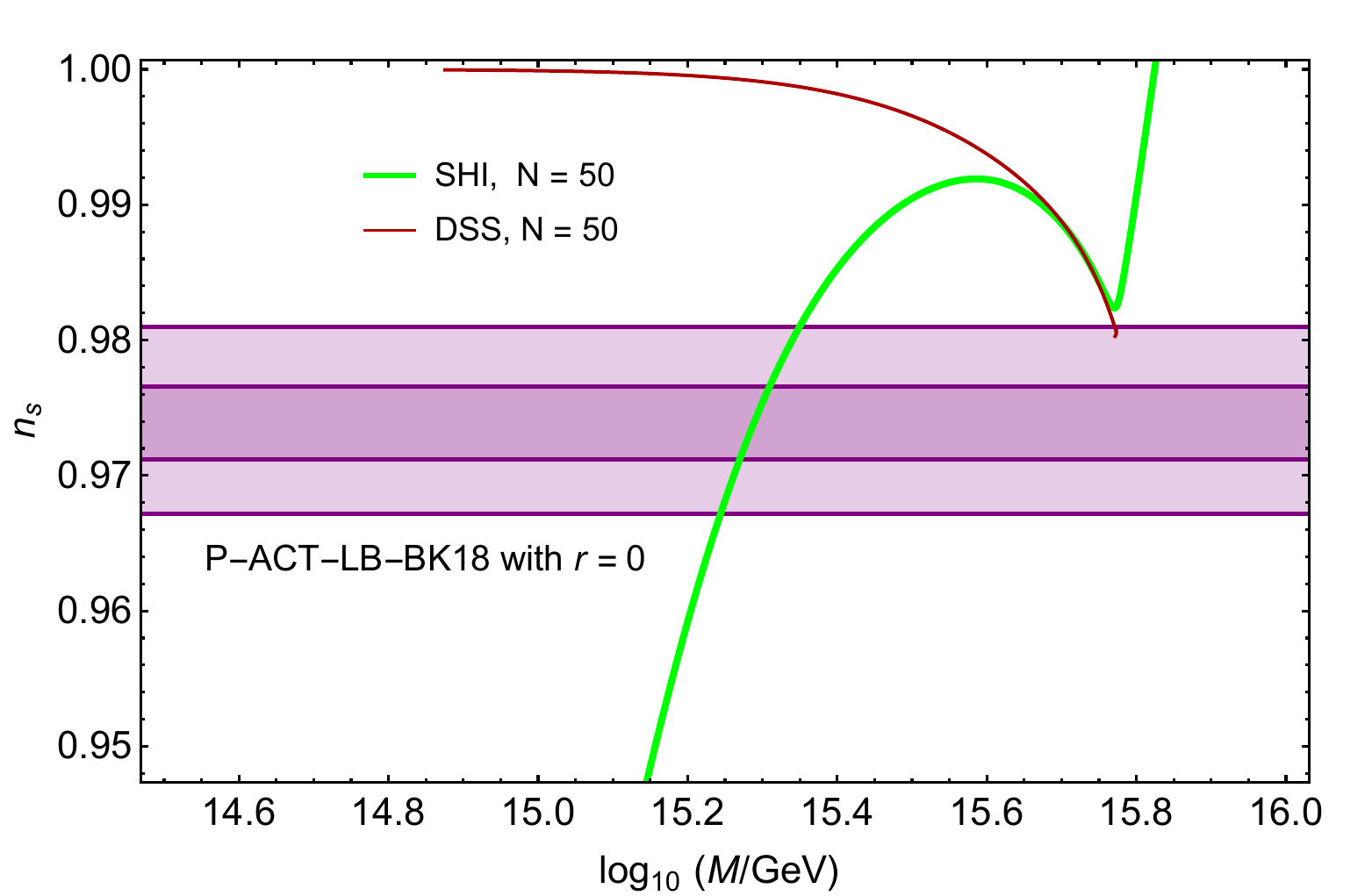}
\caption{\label{gce}$n_s$ versus $\log (M/$GeV) for SHI (which incorporates soft SUSY-breaking terms, sugra corrections, and one-loop effects), and for DSS (with one-loop corrections only), compared with ACT, Planck, BK18 [Plus LB].
}  \label{nsM}
\end{figure}

For completeness, in \cref{nsphi} and \cref{nsM}, we plot $n_s$ versus the inflaton field value and the gauge symmetry breaking scale $M$, respectively. Clearly, the field values are sub-Planckian, which is a particularly attractive feature of this minimal model. Furthermore, the gauge symmetry breaking scale is predicted to lie close to the GUT scale, as first shown in Ref.~\cite{Dvali:1994ms}.

\section{Conclusion}
Supersymmetric hybrid inflation models, referred to as spontaneously broken supersymmetry models by the Planck collaboration, are based on well-motivated extensions of the Standard Model. 
In the minimal model, defined by a renormalizable superpotential and a canonical K\"ahler potential, the gauge symmetry breaking associated with inflation is estimated to lie in vicinity of the GUT scale.
The scalar spectral index $n_s$, based on an inflationary potential that includes all relevant terms, turns out to be fully consistent with the Planck measurements as well as the more recent Atacama Cosmology Telescope observations. 
We have followed the Planck notation for the inflationary potential in these models in order to make the discussion more transparent for comparisons with all future observations.
In particular, we have included two additional contributions in the inflationary potential for the minimal supersymmetric model which were left out, presumably inadvertently, in the Planck analysis of this model.  
We have seen that the tensor-to-scalar ratio $r$ lies well below the observable range in the minimal model. However, this is easily remedied in non-minimal extensions based on realistic models where $r \lesssim 0.01$, which should be accessible in the near future \cite{Shafi:2010jr,Rehman:2010wm,Civiletti:2014bca}. 

\section{Acknowledgement}
We thank Muhammad Nadeem Ahmed and Muhammad Ibrahim for their help with the figures.

\bibliographystyle{apsrev4-1}
\bibliography{bibliography}

\end{document}